# Multiple Access Visible Light Communication using a Single Photo-Receptor

MD Rashed Rahman
*Department of Computer Science*
*Georgia State University*
Atlanta, Georgia, USA
mrahman19@student.gsu.edu

Ashwin Ashok
*Department of Computer Science*
*Georgia State University*
Atlanta, Georgia, USA
aashok@gsu.edu

*Abstract*—The directionality of optical signals provides an opportunity for efficient space reuse of optical links in visible light communication (VLC). Space reuse in VLC can enable multiple-access communication from multiple light emitting transmitters. However, traditional VLC system design using photo-receptors requires at least one receiving photodetector element for each light emitter, thus constraining VLC to always require a light emitter-to-light receiving element pair. In this paper, we propose, design and evaluate a novel architecture for VLC that can enable multiple-access reception using a photoreceptor receiver that uses only a single photodiode. The novel design includes a liquid-crystal-display (LCD) based shutter system that can be automated to control and enable selective reception of light beams from multiple transmitters. We evaluate the feasibility of multiple access on a single photodiode from two light emitting diode (LED) transmitters and the performance of the communication link using bit-error-rate (BER) and packet-error-rate (PER) metrics. Our experiment and trace based evaluation reveals the feasibility of multiple LED reception on a single photodiode and estimated throughput of the order of Mbps.

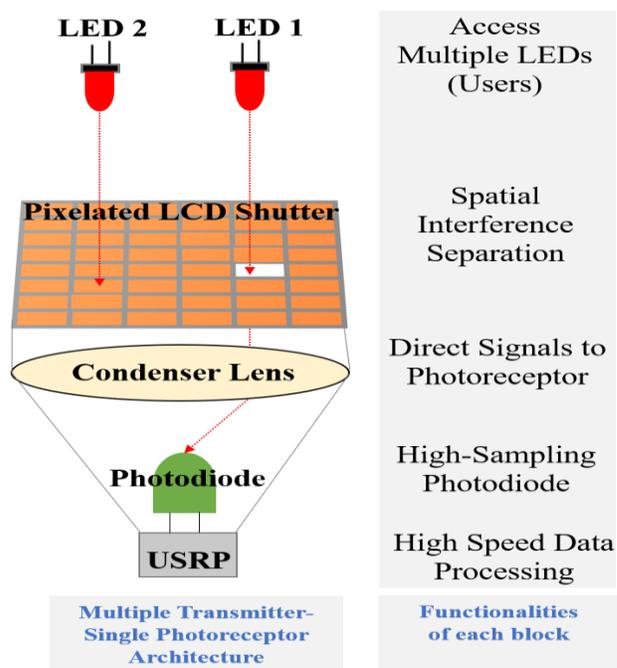

Fig. 1. Pixelated Shutter VLC Architecture for Multiple Access Reception

## I. INTRODUCTION

The significant growth in wireless data traffic has opened up new opportunities for emerging communication methodologies that use unused bands of the electromagnetic spectrum. Visible Light Communication, or *VLC* [1], [2], is an emerging wireless communication technology that operates unregulated in the visible–light band (400–800 THz frequencies or 380–780 nm wavelengths) of the electromagnetic spectrum, and is enabled by light emitting elements such as light emitting diodes (LED) and light receiving elements such as photodiodes (PD). Due to directionality of light beams, VLC is a line-of-sight (LOS) technology that requires the light transmitter and receiver to be within each others field-of-view (FOV). The LOS requirement provides novel opportunities for efficient space reuse in VLC where multiple light emitting transmissions could be multiplexed.

Traditional VLC that operates using photodiode based reception, has not been designed for receiving multiple transmissions from light emitters at the same time. While it may be possible to multiplex signals by combining space division multiple access (SDMA) with time/frequency (TDMA/FDMA) division access schemes, the nature of photoreceptors to ADD all the detected photons within its FOV limits the use of only a single light beam at each instance of time. While Multiple-Input Multiple-Output (MIMO) architectures for VLC that have been proposed and designed before [3], [4] before are clearly relevant, these architectures require multiple photoreceptor elements, for example, as in photodetector arrays and image sensing arrays or cameras. The use of array elements in a receiver limits the sampling bandwidth of the reeiver hence limiting the achievable throughput of the system. Enabling multiple-access while retaining the high-speed sampling capacity of photoreceptors has not been explored before, and is the key challenge we address through this work.

In this paper, we design a receiver design that can enable multiple light beam reception through a single photodetector using a novel architecture that uses a hardware shutter mechanism to intelligently allow/disallow signals at the receiver. We build upon our design which we had first proposed in our preliminary work [5] that introduced the *pixelated-shutter* architecture. In this work, we design a shutter control mechanism to automatically enable the selective reception of signals on the single photodiode receiver. The key issue with multiple LED transmission is the fact that the optical signal energy get added at the receiver over its sampling duration. This limits differentiating the actual signal from ambient noise or an active noise or interference source. As can be seen in the illustrative diagram of our pixelated shutter system in Fig. 1, if we consider LED 1 as the signal of interest at an instance of time, LED 2 becomes the interferer or noise to the photodiode. Our system, using the automated shutter in conjunction with a data-packetized multiple access scheme, can differentiate the signals and enable selective reception from either of the transmitters.

At a high-level, we believe that the functionality provided by our system can be treated equivalent to enabling MIMO using only single receiver (non-array receiver) using a single antenna (1 photodiode element). To the best of our knowledge, our system presents the feasibility and evaluation of such a functionality using a new architecture. In summary, the key contributions of this work are:

1) Feasibility tests of high-speed reception, through bit-error-rate (BER) analysis, using our prototype pixelated shutter system under a multi transmitter scenario (2 independent transmitters).
2) Design of an automated shutter protocol for selective reception in a multiple LED transmitter–single photodetector scenario.
3) Trace-based evaluation of automated shutter control mechanism performance for transmitter differentiation through packet-error-rate (PER) and latency metrics.

## II. BACKGROUND AND RELATED WORK

**Background.** In our prior work [5], we introduced our hybrid architecture design of VLC that uses a highspeed photodetector and an LCD shutter acting as a programmable image sensor aperture. In this work, measurement studies proved that noise and interference can be separated spatially using our VLC receiver to improve the Signal to Noise Ratio(SNR) and Signal to Noise Interference Ratio (SINR) significantly. This work leads and consolidates the idea of multiple access in VLC by adopting a shutter controlling algorithm in the receiver. Through a proof-of-concept experimentation, in our previous work [5] we have studied the feasibility of noise and interference reduction by manually selecting one of the shutter pixel apertures for higher signal reception. In this paper, we relax the assumption of the apriori knowledge of which shutter should be opened (or closed), and advance the design by providing automation and digital identification based intelligent selection of the shutter control.

**Non-Orthogonal Multiple Access in VLC.** Recent works have proposed non-orthogonal multiple access (NOMA) schemes for VLC [6], [7] to improve spectral efficiency and enable multiple access in VLC. However, the common challenge in these designs is the reliance on channel state information (CSI) and the requirement for the transmitter and receiver to be informed apriori about CSI. For example, the work in [6] introduces a power domain based multiple access protocol so that the users can use the entire bandwidth during the communication session, but requires the CSI and only works in small indoor environments. NOMA is a good contender for multiple access in VLC however the designs have been largely limited to showing the feasibility of interference cancellation under strong assumptions which limit the effective throughput performance of the VLC system.

**Multiple Access using MIMO Techniques.** Date rates in VLC can be significantly improved by multiplexing data communication across multiple LED transmitters and photodiode receivers channels. Using the MIMO technique proposed in [8], several works have presented different multiple access schemes using different equalization [9] and modulation schemes such as OFDM [10], [11]. The paper [12], introduces an Optical Code Division Multiple Access (OCDMA) technique and the work [13] used intensity modulation to support multiple users in MIMO VLC system. A key challenge with photodiode arrays is that they allow more noise, from ambient light (sunlight and artificial lighting), into the receiver due to the wide field–of–view, thus affecting received signal quality and data rate. In the work from [14], the authors present a precoding technique to mitigate inter-cell and intra-cell ambient light interferences in multi-cell VLC systems to improve the bandwidth efficiency, where specific spatial regions were considered as cells, similar to cellular communication. However, the complexity of the MIMO techniques for real-time implementation and performance, and the necessity of high efficiency and costly photoreceptors (avalanche photodiodes [4]) for improving data rates, have kept MIMO approaches still in their early stage for adoption in VLC.

**Wavelength Division Multiplexing (WDM).** The idea of using WDM to provide multiple access in VLC is another key trend in the research community. In paper [15], the authors introduce a bi-directional VLC in full duplex mode by parallel transmission of three (RGB) channels and use OFDM modulation demodulation to increase the aggregate date rate. However, this design incurs a high bit error rate. In another work [16], a color-shift keying CDMA (CSK-CDMA) based VLC system has been developed to increase the VLC throughput and for allowing multiple access. While this system can avoid interference across specific wavelengths, the complexity and cost limit the usage of these approaches due to requirement of array receivers or customizations such as usage of specific light emitting wavelengths.

## III. SYSTEM ARCHITECTURE

We have designed a pixelated shutter based VLC receiver that automatically and intelligently identifies and selects/isolates signals from multiple LEDs. The key components of the system, as illustrated in Fig. 1, include a photodiode (capable of high-speed sampling), a pixelated LCD shutter, a shutter control unit, a computing unit, and a condenser lens for optical focusing. The key idea of proposed design is to allow signals from multiple LED transmitters to be correctly detected and decoded using a single photodetector VLC receiver, while not compromising the data rate of the system.

The shutter uses LCDs which act as a digital aperture that allows (disallows) the impinging light beams, to reach the photodiode, based on the input voltage to the shutter. Using this digital aperture as a control the receiver is able to select which of the incoming light beams are to be decoded by the photodiode at each instance of time. The computing unit at the receiver hosts the decoding algorithms and mechanisms to efficiently decode the signal that has been selected. The digital control of the shutter is integrated with the decoding modules in the computing unit, such that there is active feedback on the quality of the received signal. The feedback information includes the received signal-to-noise-ratio (SNR) and a digital identification of the signal using packet header bits. This design enables a seamless functioning of the selective control of the reception and the decoding in tandem. The selection of the desired signal is a one-time process and needs to be repeated only when there is mobility or during link failures.

*A. Spatial Multiplexing using Pixelated Shutter*

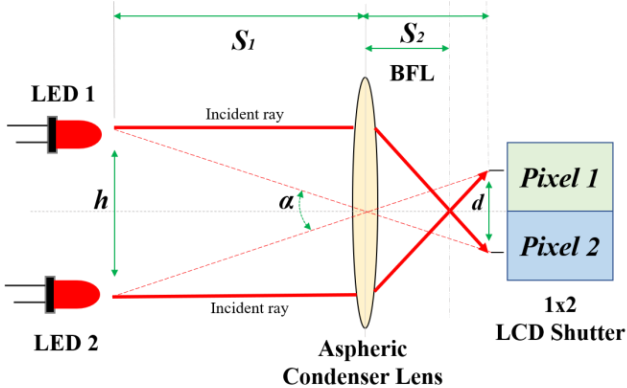

Fig. 2. LED Positioning to focus on separate pixels

We describe the spatial multiplexing setup through the illustration provided in Fig. 2. Let us consider two LEDs placed in space (at same height but separated along the horizontal) aiming to focus their signals (using the condenser lens) onto a photodetector (not shown in Fig. 2) by passing through a 1 x 2 LCD shutter pixel system, where the pixels are aligned next to each other along the horizontal. Let us consider that each pixel $i$ is responsible for signals from corresponding LED $i$. In this way, when the signals from the LEDs are beamed onto the photodiode, each pixel can selectively allow/disallow the signals provided the signals are independently identified (and differentiated) and the information on which signal (LED) should be selected is feedback to the shutter control unit (not shown in Fig. 2).

We make an assumption that each LED signal can be spatially separated onto specific pixels independently. Through lens equation [17] and using simple trigonometrical calculations, we derive that the minimum distance of separation between the LEDs must be $h = dS_1/BFL$ where $d$ is the distance between two pixel centers (considering a square pixel it is the pixel side length), $S_1$ is the distance from LED to the lens, and $BFL$ is the back focal length which is the focal length of the lens. We also derive the minimum angle of separation between the LED beams as $\alpha = 2 \arctan(h/2S_1)$.

We justify the assumption by calculating the practical minimum distance of separation to be maintained between the LEDs to ensure non-overlap on a pixel. For the explanation we will use values from the prototype setup in our laboratory, where $S_1 = 15.5cm$, $S_2 = 8.2cm$ and $BFL = 3.75cm$ and $d = 3.6cm$. Based on these values, the minimum horizontal separation between the LEDs has to be $h = 14.88cm$ and $\alpha = 51.2°$. These equations provide the designer the control of placing the LEDs in space so as to allow for multiplexing using our system. For example, at a distance of 10m the equivalent minimum separation distance would be close to 10m, however, this distance can be reduced if the $d$ were to be increased. By merely doubling the size ($d$) of the shutter, the required distance $h$ now can be 5m. This means that the selection of separation between LEDs in space and the size of the shutter involves a tradeoff. Note that these calculations are considering the radiation angle of the LED is of the order of $50°$. The tradeoff between $h$ and $d$ can also be adjusted when having a smaller radiation angle LEDs, for example LASER type emitters.

*B. Automated Shutter Control Protocol Design*

The LCD shutter does the selection of what transmit signal to receive and what to filter out using an automated shutter control protocol that we have developed. The shutter control protocol operates in tandem with the decoding process running in the computing unit connected to the photodiode. The control protocol, discussed in Algorithm 1 involves two steps:

**Step 1 or the Discovery phase**, where the receiver does a preliminary pruning of all signals that do not represent a transmit signal. For example, ambient light, DC noise sources and known noise sources are eliminated. This way, only a subset of the shutter pixel array are kept OPEN and are to be processed, thus limiting the processing to a smaller subset of signals.

**Step 2 or the Identification phase**, where the receiver does a fine tuning of identifying each transmit signals and selectively opening the corresponding shutter pixels to allow/disallow the signal for continued reception at the photodiode. The identify of the signals are maintained through unique header sequences in the data packets (e.g. barker codes). We consider that a unique ID of each transmitter will be registered at the receiver apriori during the first setup of the system (one time) and update the ID look-up table as necessary.

The shutter control protocol enables multiple access where information from multiple LED transmitters to be decoded by a single photodiode receiver. By enabling which signals to receive at which instance of time, the receiver can choose time-slots to receive and decode specific signals. Our system by default functions as a space and time-division multiple access system where each transmission is decoded across a time-slot duration of $Ts$ seconds. The choice of the value of $Ts$ depends on the application. For example, a slot duration of 1-2 seconds may work for beaconing and repetitive transmission such as in sensor or IoT applications, however, for streaming applications the slot has to be made much smaller (order of few ms). A smaller slot duration also implies that the pixel switching control must happen as fast as the selected slot duration. Depending on the type of LCD shutters, the switching time can vary from few micro to 10s of milli seconds.

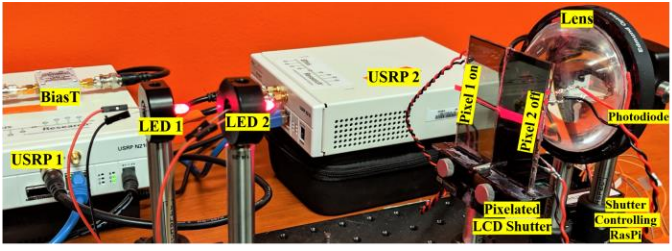

Fig. 3. General setup of the pixelated shutter receiver system. This picture shows a 2 LED transmitter setup with a single photodiode receiver and 2 LCDs fit in a 1 x 2 (pixels) configuration.

**Algorithm 1** Automated Shutter Control Protocol

**Result:** Allow desired optical signals and block interference
**Initialization:**
 Open all pixels (px1, px2, px3. ...pxn) to allow signals
 Set a fixed switching time (Ts) for all pxi (i=1,2. N)
 Preset an empirical threshold SNR value (SNRth)
 Refresh unique identifier (ID) look-up table
**Step 1: Discovery**
 Iterate each pixel OPEN and all others CLOSED for duration Ts and record SNR
**if** *(SNR(pxi)) ≥ SNRth* **then**
 | (i) OPEN pxi pixels.
 | (ii) Shut off the remaining pixels
 | (iii) Proceed to Step 2 (Identification)
**else**
 | CLOSE all the pixels and refresh the program
**end**
**Step 2: Identification**
 Correlate decoded signal ID with look-up table IDs
**if** *ID matches* **then**
 | (i) OPEN only matched ID containing pixels
 | (ii) CLOSE all the remaining pixels.
**else**
 | CLOSE all the pixels and goTo Step 1
**end**

## IV. PROTOTYPE IMPLEMENTATION

We implemented the automated shutter control mechanism on our preliminary pixelated shutter testbed as shown in the setup in Fig.3. The key components of the system include 2 RED LEDs [18], a PDA10A2 Amplified Photodetector [19], a custom made 1x2 pixelated LCD shutter and an aspheric condenser lens (outer diameter 80 mm and $BFL = 37.5mm$) [20]. A Raspberry Pi 3 Model B+ [21] hosts the shutter controlling algorithm and interfaces with a 1x2 pixelated shutter, which we built using off-the-shelf LCD shutter elements from Adafruit [22]. We used two N210 USRPs [23] as the computing units at the transmitter (controlling LED transmissions) and receiver ends (decoding signals from photodiode). We used a LFTX daughterboard [24] capable of operation from 0-30MHz and a RFTX daughterboard [25]. The 2 LEDs were controlled using two different USRPs, each hosting a LFTX, and one of the USRPs hosting a LFRX that also conducted the reception. We used GNU Radio blocks [26] (block diagram of GNU applications shown in Fig. 4) to transmit and receive signals using the USRPs. We chose to use the state-of-the-art Gaussian Minimum Shift Keying (GMSK) as the modulation strategy in our design, however, any type of modulation could be used in the system. As shown in Fig. 5, we used a 13-bit and 11-bit barker sequences for LED1 and LED2, respectively. We have implemented the transmissions in the form of UDP packets of size 2096 bits, which equals the packet size of a typical (proposed) IEEE 802.15.7 VLC standard [27] packet size.

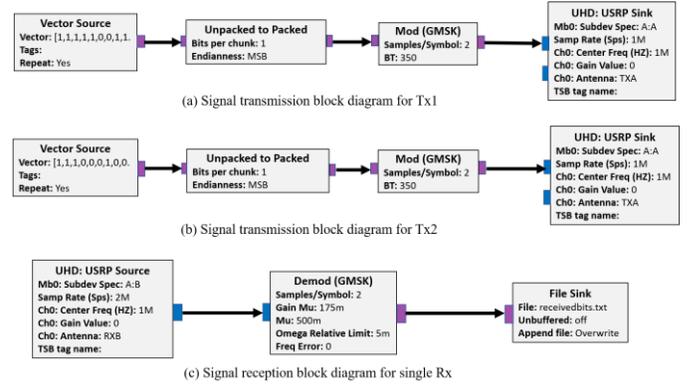

Fig. 4. Illustration of USRP applications' GNU Radio Block Diagram for transmitters (1 and 2) and the single receiver. The vector sources change for different experiments.

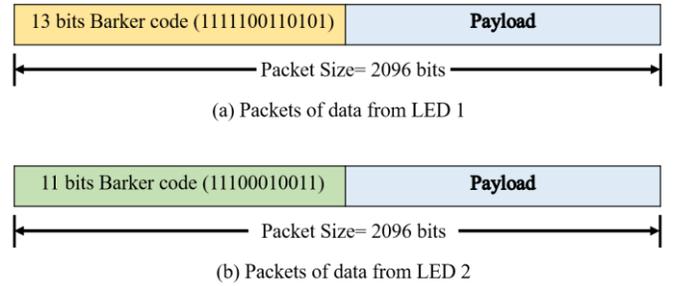

Fig. 5. Transmitted Packets from LED 1 and LED 2

## V. EVALUATION

We evaluate our system to study the feasibility of our system to achieve high speed reception and the performance of the automated shutter protocol for multi channel visible light signal reception on a single photodiode receiver. All the experiments were conducted in a lab setting, indoors, under ceiling white ambient lighting. Unless mentioned, the distance between the LED and photodiode in our experiments was set to 15.5cm.

### A. BER Analysis

We conducted two types of experiments to evaluate the bit error rate performance of our system: (a) BER under different types of interference, and (b) BER under selective mapping of signal and shutter pixels. The BER experiments involved, in general, transmitting a random stream of $10^4$ bits in a loop and logging the decoded bits at the receiver. We chose LED 1 as the desired transmitter and LED 2 as interference. The frequency of communication was chosen as per the experiment goals.

*1) BER and Interference Patterns:* We set the transmit frequency to be 100Hz and conducted the BER experiment under four signaling types:

**Type 1:** Only the transmit signal

**Type 2:** The transmit signal and ambient DC noise

**Type 3:** Transmit signal and Interference signal sending identical patterns in phase

**Type 4:** Transmit signal and Interference signal sending identical patterns at 180 deg out-of-phase.

| Type 1 | | Type 2 | | Type 3 | | Type 4 | |
|---|---|---|---|---|---|---|---|
| Case I | Case II | Case I | Case II | Case I | Case II | Case I | Case II |
| $10^{-3}$ | $10^{-3}$ | $4.9*10^{-1}$ | $10^{-3}$ | $10^{-3}$ | $10^{-3}$ | $4.9*10^{-1}$ | $10^{-3}$ |

TABLE I
BER AT 100 HZ SIGNALING UNDER DIFFERENT INTERFERENCE PATTERNS. CASE 1: ALL PIXELS OPEN, CASE 2: ONLY DESIRED SIGNAL PIXEL IS OPEN.

We report the BER from these experiments in Table I. We can observe from the BER values from Case 1 that the error rates for the system are generally high when the desired signals and the interference are combined at the photodetector. We observe that the BER is practically low (for feasible data communication) when the pixels are selectively OPEN/CLOSE to allow only the desired signal, which has been achieved without major changes to the receiver. We can also observe from the BER values, the additive property of the receiver, where the BER is low when the interference signal is identical and of same phase (as the effective received signal amplitude is doubled) and high when the same interfering signal is out-of-phase. The 100Hz frequency was chosen arbitrarily, as the primarily goal of this experiment is to validate the additive nature of optical signal at the photodiode receiver.

*2) BER and Selective Signaling:* Consider pixel-1 as the pixel corresponding to LED 1 and pixel-2 as the one for LED 2. We conducted the BER experiments under three different configurations of the shutter pixel and under three different transmit frequencies. The results from Table II indicate that the BER is at least an order low when only the desired signal is received versus when the interference is also sampled on the single photodiode receiver. The BER values, though relatively high (which can be reduced using error control coding) for data streaming applications, however, indicate feasibility of multiple access using our proposed architecture. Considering, *Goodput ≈(1-BER)\*coderate\*transmitsymbols/sec\*bits/symbol*, we note that at a hypothetical error control code rate of 1/3, 2Mhz (2M symbols/sec) transmit rate and 2bits/symbol modulation rate, the effective achievable *Goodput* in our preliminary system can be estimated as 1.3 Mbps.

| Frequency (Hz) | Config.1 | Config.2 | Config.3 |
|---|---|---|---|
| 500 KHz | 0.015 | 0.039 | 0.21 |
| 1 MHz | 0.015 | 0.035 | 0.21 |
| 2 MHz | 0.015 | 0.030 | 0.21 |

TABLE II
BER AT DIFFERENT SHUTTER PIXEL CONFIGURATIONS. **Config.1:** ONLY PIXEL-1 IS OPEN. **Config.2:** ONLY PIXEL-2 IS OPEN. **Config.3:** BOTH, PIXEL-1 AND PIXEL-2 ARE OPEN.

### B. PER and Signal Selection

| Freq(Hz) | # Pkts in $Ts$ | Packet Error Rate (PER) (%) | |
|---|---|---|---|
| | | **Pixel 1** | **Pixel 2** |
| 500 KHz | 477 | 5.88 % | 5.46 % |
| 1 MHz | 954 | 4.83 % | 2.63 % |
| 2 MHz | 1908 | 3.36 % | 3.25 % |

TABLE III
PER CALCULATED FROM RECEIVED SIGNAL TRACES FROM EACH PIXEL OPEN DURATION OF $Ts$ = 2 SEC. PIXEL 1(2) CORRESPONDS TO SIGNAL FROM LED 1(2).

We transmitted a continuous stream of packets of size 2096 bits, where each packet had a random stream of bits as payload and each packet header was of size 13 bits. These served as the unique IDs for the transmissions from the LEDs. The header would be the same for all packets from a specific transmitter. For LED 1 we chose a 13 bits sampled from barker sequences as ID, and for LED 2 we chose a 11 bits sampled from barker sequence plus a 2 bit (11) padding. We chose the time slot ($Ts$) duration to be 2 seconds for each iteration over a pixel. We recorded the received and decoded bits from packets in each iteration of pixel OPEN cycles. We collected the received signal traces from each iteration of each STEP of the automated shutter control protocol (algorithm 1). We iterate over 1 cycle of each pixel OPENING and then choosing either of the pixels that corresponds to the desired signal. We conducted 3 trials each for considering LED 1 (2) as the desired signals and determined the PER through offline calculations using the traces.

## C. Processing Latency

| Pixels | Pixel 1 | Pixel 2 | Switching time [ms] |
|---|---|---|---|
| $SNR_{db}$ | 19.97 | -0.27 | 100 ms |
| | 19.96 | 0.48 | 500 ms |
| | 19.86 | -0.47 | 1000 ms |
| | 20.02 | -0.60 | 1500 ms |
| | 19.99 | -1.18 | 2000 ms |

TABLE IV
SNR (dB) FOR EACH PIXEL OPEN UNDER DIFFERENT SWITCHING TIMES.

As the switching time (or time slot duration $Ts$) can impact the signal quality from each pixel during the shutter control phases, we first validated the consistency of the SNR values under different switching duration of the pixels. Considering LED 1 as desired signal and with LED 2 switched OFF, we alternated pixels 1 and 2 to be OPEN for specific duration and recorded signal and noise power. We report the SNR values in Table IV and observe the consistency of SNR for short (100ms) as well as long switching durations (2sec).

| Resolution (pixels) | Step 1 [ms] | Step 2 [ms] | Total Time [ms] |
|---|---|---|---|
| 100x100 | 10 | 100*2096*1µs=209.6 | 219.6 |
| 1000x1000 | 1000 | 100*2096*1µs=209.6 | 1209.6 |

TABLE V
ESTIMATED PROCESSING LATENCY IN EACH STEP DUE TO AUTOMATED SHUTTER CONTROL UNDER DIFFERENT SHUTTER RESOLUTIONS.

Let us consider switching times of $Ts$ = 1µs, and shutter resolutions of 100x100 and 1000x1000 pixels. Considering a hypothetical number of 100 effective transmitters, which map to 100 different pixels, the effective, theoretically estimated, processing time of Steps 1 and 2 in Algorithm 1 would be as reported in Table V, and can be observed to be within practical working latency for a wide range of VLC applications including sensing, IoT and low-speed device-device data transfers.

## VI. CONCLUSION AND FUTURE WORK

In this paper, we introduced a novel architecture and a protocol to enable multiple access reception on a VLC receiver with only a single photodiode element. We designed and evaluated a system that enables transmission from 2 LEDs simultaneously and selectively decodes packets from each based on a selection algorithm that uses OPEN/CLOSE cycles of LCD shutter pixels acting as digital apertures for the photodiode signal. Through BER, PER and latency metrics, computed through experiments, we showed the feasibility and performance of our 2 transmitter-to-1 receiver multiple access system at low and high signal frequencies. To the best of our knowledge, this work sets the foundation stage for future work in multiple access using single photodiode receivers.